\documentclass[12pt,preprint]{aastex}
\usepackage[usenames]{color}

\shorttitle{Solar Flare Prediction Model with Machine-Learning Algorithms}
\shortauthors{N. Nishizuka et al.}

\begin{document}

\title{SOLAR FLARE PREDICTION MODEL WITH THREE MACHINE-LEARNING ALGORITHMS USING ULTRAVIOLET BRIGHTENING 
AND VECTOR MAGNETOGRAM}


\author{N. Nishizuka\altaffilmark{1}, K. Sugiura\altaffilmark{2}, Y. Kubo\altaffilmark{1}, M. Den\altaffilmark{1}, S. Watari\altaffilmark{1} and M. Ishii\altaffilmark{1}}
\altaffiltext{1}{Applied Electromagnetic Research Institute, National Institute of Information and Communications Technology, 4-2-1, Nukui-Kitamachi, Koganei, 
Tokyo 184-8795, Japan; nishizuka.naoto@nict.go.jp}
\altaffiltext{2}{Advanced Speech Translation Research and Development Promotion Center, National Institute of Information and Communications Technology}

\begin{abstract}
We developed a flare prediction model using machine learning, which is optimized to predict the maximum class of flares occurring in the following 24 h. Machine 
learning is used to devise algorithms that can learn from and make decisions on a huge amount of data. We used solar observation data during the period 2010 - 
2015, such as vector magnetogram, ultraviolet (UV) emission, and soft X-ray emission taken by the Solar Dynamics Observatory and the Geostationary Operational 
Environmental Satellite. We detected active regions from the full-disk magnetogram, from which $\sim$60 features were extracted with their time differentials, 
including magnetic neutral lines, the current helicity, the UV brightening, and the flare history. After standardizing the feature database, we fully shuffled and randomly 
separated it into two for training and testing. To investigate which algorithm is best for flare prediction, we compared three machine learning algorithms: the support 
vector machine (SVM), k-nearest neighbors (k-NN), and extremely randomized trees (ERT). The prediction score, the true skill statistic (TSS), was higher than 
0.9 with a fully shuffled dataset, which is higher than that for human forecasts. It was found that k-NN has the highest performance among the three algorithms. 
The ranking of the feature importance showed that the previous flare activity is most effective, followed by the length of magnetic neutral lines, the unsigned 
magnetic flux, the area of UV brightening, and the time differentials of features over 24 h, all of which are strongly correlated with the flux emergence dynamics 
in an active region.
\end{abstract}


%
\section{Introduction}

The mechanism of solar flares is a long-standing puzzle in solar physics. The energy storage and triggering processes of flares are driven by the emergence of flux in 
the photosphere \citep[e.g.,][]{pri02, shi11,tak15}, and flares are directly observed by a photospheric magnetogram. The shape and complexity of sunspots in white-light 
emission have been classified according to the sunspot growth level \citep{mci90}. It is empirically known that larger sunspots with a large number of umbra and a 
more complicated magnetic flux structure tend to produce larger flares \citep[e.g.,][]{sam00, gal02, col09, li08, blo12, lee12, bar16}, as well as repeated flares in 
the same active regions (ARs) \citep[e.g.][]{zir88, zir91}.

Features derived from the line-of-site magnetogram are useful indicators for future flare prediction, such as the magnetic flux, the gradient of the magnetic field 
\citep{yu09, ste11}, the length of magnetic neutral lines \citep{ste11}, the effective magnetic field \citep{geo07, pap15}, the unsigned magnetic flux near the magnetic 
neutral lines \citep[R-value:][]{sch07, fal11}, the total magnetic energy dissipation \citep{son09}, the weighted magnetic neutral line length and the distance between 
NS polarity sunspot centers \citep{mas10}, the non-potentiality \citep[e.g.,][]{fal14}, and the wavelet spectra \citep{yu10, alg15, bou15, mur15}. These features are 
related to the dynamics of flux emergence and are strongly correlated with the energy storage and the triggering mechanisms.

Leka \& Barnes (2003) pioneered the use of vector magnetic field data for flare prediction, and the features from a vector magnetogram were first used with machine 
learning by Bobra \& Couvidat (2015). Detailed vector magnetogram observations show the dynamic variation of the magnetic configuration near magnetic neutral lines 
caused by successive flux emergence \citep{kub07}, and the photospheric flow around magnetic neutral lines has been shown to be an important indicator of the 
occurrence of flares \citep{wel09}. Recently, a model of flare triggers has been proposed by Kusano et al. (2012), in which the relative direction of an emerging flux 
near magnetic neutral lines to the pre-existing sheared magnetic loops determines the size of flares; this model has been supported by observations \citep{bam13, tor13}.

As an emerging flux appears near magnetic neutral lines in an AR, small-scale energy release occurs in the lower chromosphere via magnetic reconnection, which 
has been observed using the 1600 \,\AA\ filtergram of the Transition Region and Coronal Explorer (TRACE) as a gradual increase in the ultraviolet (UV) emission 
for 2-3 h in the preflare phase \citep{sai07}, as well as in Ca$_{\rm II}$ H emission by Hinode observations \citep{bam13}. The 1600 \,\AA\ filtergram observes the 
UV continuum, chromospheric lines, and the C$_{\rm IV}$ doublet ($\sim$1550 \AA), which is strongly enhanced and well correlated with the hard X-ray emission 
\citep{bre96, han99, war01, nis09}. Moon et al. (2004) found UV brightening at one end of a pre-erupting filament, where magnetic reconnection occurs in the 
low atmosphere and changes the magnetic connectivity, leading to the initiation of the filament eruption \citep[see also][]{kim08, guo12}.

The amount of recent open-access solar observation data is so large that it is beyond human processing ability. To deal with the data, several machine-learning 
algorithms \citep[see an introductory text to machine-learning, e.g.][]{has09} have been applied to the flare prediction problem: a neural network \citep{qah07, col09, 
hig11, ahm13}, C4.5 decision trees \citep{yu09, yu10}, learning vector quantization \citep{yu09, ron11}, a regression model \citep{lee07, son09}, k-nearest neighbor 
\citep{li08, hua13, win15}, a support vector machine \citep{qah07, bob15, mur15}, a relevant vector machine \citep{alg15}, support vector machine regression 
\citep{bou15}, and an ensemble of four predictors \citep{gue15}. However, the best algorithm for flare prediction was not discussed in the previous works, and 
it cannot be found without directly comparing the performances of different algorithms.

Thus, in this paper, we compared three machine-learning algorithms to find which algorithm has the highest performance for a flare prediction. We also extended 
the observation data period and wavelength obtained by the Solar Dynamics Observatory \citep[SDO;][]{pes12} and optimized each algorithm to improve the prediction 
accuracy. Novel features such as UV brightening and the vector magnetogram have been included, and finally the importance of different features was calculated and 
ranked. In section 2, we give an overview of our prediction model, which is explained in detail in section 3. The prediction results are described in section 4 and a 
discussion and conclusion are given in section 5.

\clearpage
%
%
\section{Overview of our Prediction Model}

The procedures of our flare prediction model are as follows, (i) First, observation data are downloaded from the web archives of SDO and the Geostationary Operational 
Environmental Satellite (GOES), such as the line-of-sight magnetogram, vector magnetogram, 1600 \,\AA\ broadband filtergram images, and the light curves of the soft 
X-ray emission. (ii) Second, ARs are detected from full-disk images of the line-of-sight magnetogram, and the ARs are tracked using their time evolution. (iii) For each 
AR, features are calculated from multiwavelength observations, and flare labels are attached to the solar feature database if an X/M-class flare occurs within 24 h after 
an image. (iv) Supervised machine learning is carried out with a 1 h cadence to predict the maximum class of flares occurring in the following 24 h.

Our observation data are from June 2010 to December 2015, which were taken by SDO, launched in February 2010. During this period, 29 X-class and 433 M-class 
flares were observed on the disk, accounting for 90\% of the flares observed during the period. The other 10\% of the flares occurred on the limb and were removed 
from our event list. We call the events of data with flare labels ''positive events'', while the other events are ''negative events''. It is considered that X-class flares occur 
an average of 5-10 times per year during the solar maximum period, while M-class flares occur 100 times per year. Negative events are much more common than positive 
events, making the flare prediction problem an imbalanced problem.

We used the line-of-sight magnetogram taken by the Helioseismic and Magnetic Imager \citep[HMI;][]{sche12} on board SDO, as well as the vector magnetogram. The 
UV continuum of the lower chromosphere was taken by the 1600 \,\AA\ broadband filtergram of the Atmospheric Imaging Assembly \citep[AIA;][]{lem12} on board SDO. 
The full-disk integrated X-ray emission over the range of 1-8 \,\AA\ was observed by GOES. The time cadence of the line-of-sight magnetogram is 45 s, that of the 
vector magnetogram is 12 min, that of the 1600 \,\AA\ filtergram is 12 s, and that of GOES is less than 1 min. Thus, the total size of the observation dataset is so large 
that we reduced the cadence to 1 hour, in accordance with the forecast operation. The vector magnetogram data consist of the absolute field strength, the inclination 
angle, the azimuth angle, and the sign to solve the 180$^{\circ}$ ambiguity problem. By converting these components to Cartesian coordinates, we calculated the features 
listed in Table 1.

\clearpage 
%
%
\section{Details of our Prediction Model}
\subsection{Detection of ARs}

First, we detect ARs to extract solar features from the images of the downloaded observation database. We used $\sim$10$^5$ full-disk images of the line-of-sight magnetogram 
for detection with a reduced cadence of 1 h (Fig. 1). The line-of-sight magnetogram was selected for AR detection because it is less noisy than the vector magnetogram and more 
suitable for the processing carried out for detection. After determining ARs in magnetogram images, the frame coordinates of the ARs were applied to other images with different 
wavelengths (Fig. 2).

Here we defined the detection rules as follows: (i) First, we smoothed the data with 64 (=8$\times$8) binning and detected the image pixels where the absolute magnetic field 
strength is larger than a threshold value, i.e., B$_{th}$=140 $G$, for convenience (Fig. 1(b)). We set the maximum value of the observation errors for the detection threshold to 
detect faint ARs. (ii) Secondly, we placed the detected pixels in squares with a side of 160 pixels ($\sim$80'') (Fig. 1(c)). Such a 80''$\times$80'' square is the minimum unit 
of the detection region. (iii) Thirdly, if two neighboring squares overlapped, they were combined to form a larger square containing two detected points. The repetition of this 
process resulted in a single large square covering the whole AR and reduced the number of detected regions (Fig. 1(d) and Fig. 2(a)).

Next we neglected ARs detected on the limb, where the magnetic structure is difficult to see owing to the projection effect and is partially hidden by the limb. Additionally, 
the quality of the data from the vector magnetogram is poor near the limb. This is why previous papers focused on the disk-center dataset. On the other hand, in an operational 
setting, it is necessary to deal with ARs near the limb to make predictions, but there have been no attempts to verify the effectiveness of a near-limb dataset. Including the 
near-limb data, the size of the database is increased, making machine learning more appropriate. In this paper, to investigate the effect of the detection regions on the prediction 
score, we compared the following three cases by including or excluding the near-limb region: the full-disk case, an intermediate case with focusing within $\pm$53$^{\circ}$ 
of the center meridian (CM) (within 4/5 of the solar radius) and the disk-center case with focusing within $\pm$37$^{\circ}$ (within 3/5 of the solar radius). Referring to Bobra 
\& Couvidat (2015), the authors only considered flares within $\pm$68$^{\circ}$ of the CM.

Furthermore, we tracked ARs moving along the axis of solar rotation and numbered them with IDs for identification. In the case of overlapping ARs in two successive images, 
we numbered them with the same ID. We detected a total of 11700 ARs from the full-disk images during the period 2010-2015. Here, we determined regions containing 
magnetic fields larger than 140 G as ARs, and our definition is different from NOAA's. Our model includes faint quasi ARs, so as not to miss even small flares occurring 
outside of NOAA's regions. If we enlarge the detection threshold, the flare occurrence rate of the dataset will increase, but we cannot avoid missing some flares. 
Furthermore, a strong magnetic field is localized, so that not the whole area of ARs may be covered by 80''$\times$80'' squares.

\begin{figure}[hbtp]
\epsscale{.62}
\plotone{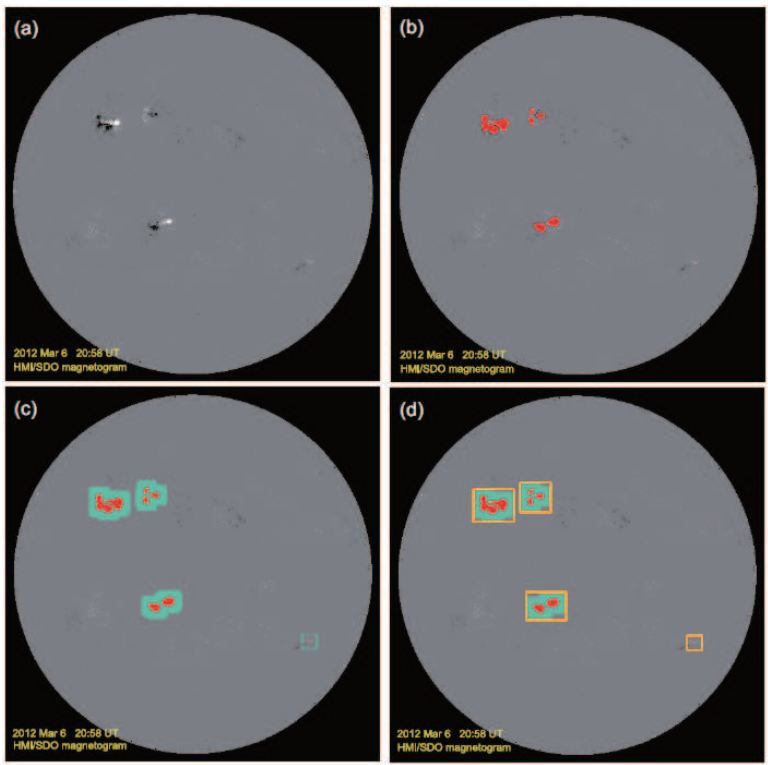}
\caption{Four steps of AR detection: (a) a full-disk magnetogram taken by HMI/SDO, (b) detected points of magnetic field larger than 140 $G$ (in red), 
(c) 80''$\times$80'' squares centering at the detected points (in light blue), and (d) coupled squares covering the whole ARs (in yellow). \label{fig1}}
\end{figure}

\begin{figure}[hbtp]
\epsscale{.62}
\plotone{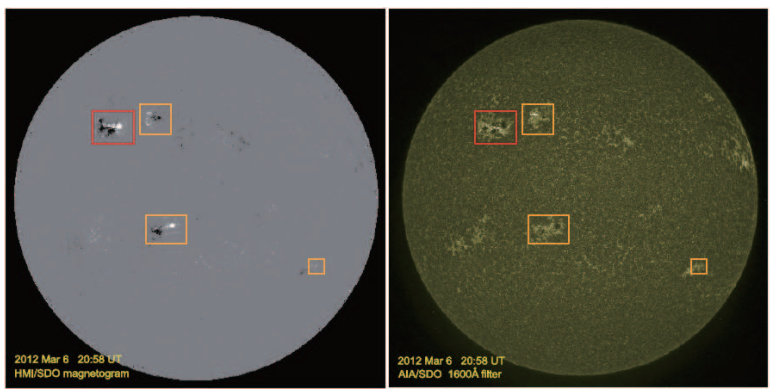}
\caption{Full-disk images of (a) the line-of-sight magnetogram taken by HMI/SDO, with detected active regions framed in yellow or red, and of (b) the UV continuum 
taken with the 1600 \,\AA\ broadband filter of AIA/SDO. The region with a red frame produced X 5.4 flare, 3 h after this image was taken. \label{fig2}}
\end{figure}

\begin{figure}[hbtp]
\epsscale{0.62}
\plotone{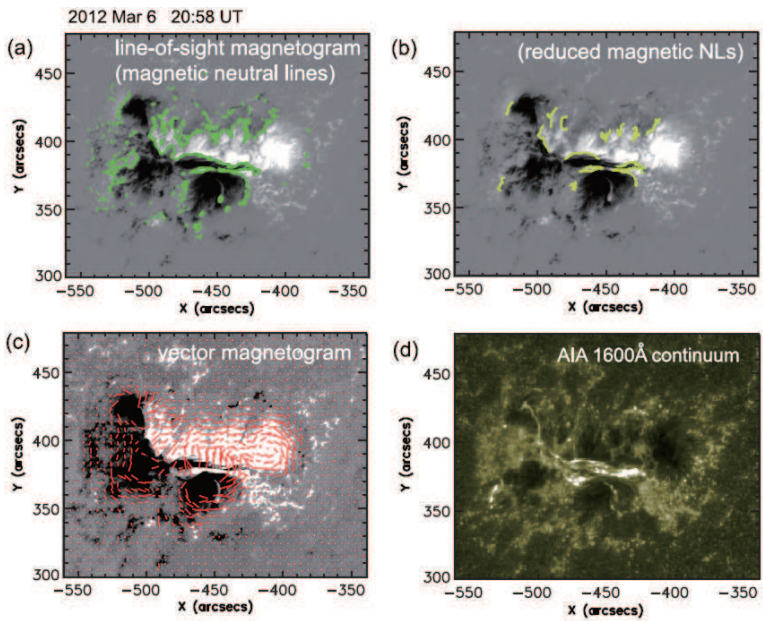}
\caption{Snapshot images of an AR, which produced X5.4 flare on 2012 Mar 7, with different wavelength observations: (a) line-of-sight magnetogram taken 
by HMI/SDO, with magnetic neutral lines in green, (b) one with only long magnetic neutral lines in yellow, (c) vector magnetogram taken by HMI/SDO, and 
(d) UV continuum using the 1600 \,\AA\ broadband filter of AIA/SDO. \label{fig3}}
\end{figure}

\clearpage
%
%
\subsection{Extraction of Solar Features}

Using the database of detected ARs, we next extracted solar features from each AR. We adopted solar features used in the previous papers, which were extracted from the 
line-of-sight magnetogram \citep[e.g.,][]{ste11, ahm13}, the vector magnetogram \citep{lek03, bob15}, and GOES X-ray data. Furthermore, in this study, we extracted the 
feature of chromospheric brightening, which was obtained from the UV continuum taken by the SDO/AIA 1600 \,\AA\ filtergram for the first time. The extracted features 
are summarized in Table 1, along with their importance ranking (which will be explained in a later section).

From the line-of-sight magnetogram, we extracted features suh as the area of an AR, the maximum $B_{LOS}$, the average $B_{LOS}$, the unsigned magnetic flux, the 
gradients of the magnetic field in the longitudinal/latitudinal directions, and the number of magnetic neutral lines. The magnetic neutral line is an indicator of flare activity 
because it is directly related to the energy storage and triggering mechanisms. We counted the number of neutral lines in an AR and measured the maximum/total length of 
the lines (Figs. 3(a)-3(b)). We detected neutral lines using two conditions: a large magnetic field gradient and a reverse of the magnetic polarity across the lines. Here we 
focused on the magnetic neutral lines longer than 100 pixels ($\sim$50'') to eliminate short and complicated neutral lines (Fig. 3(b)).

After preprocessing of the vector magnetogram, we calculated features using the three vector components of the magnetic field (Fig. 3(c)). Using the formulae in Bobra \& 
Couvidat (2015), we extracted vector magnetogram features: such as the vertical current, the current helicity, the Lorentz force, and the mean gradient of the total field. 
The formulae of the features derived from the vector magnetogram are summarized in Table 2. The corresponding features in Table 1 are marked with daggers. Moreover, 
we differentiated the extracted features with respect to time. The time derivatives of the features over 24 h, 12 h and 2 h were calculated to track the variability of ARs 
over different time scales. 

Brightening in the lower chromosphere is another indicator of flares. A few hours before a flare onset, the lower chromosphere is gradually heated, emitting light in the UV 
range (Fig. 3(d)). The brightening is located around magnetic neutral lines. We extracted chromospheric (UV) features and used them for training in the machine learning for 
the first time; these features included the maximum intensity, the brightening area, and the total intensity of UV brightening in a whole AR. We used AIA 1600 \,\AA\ 
filtergram images of SDO representing the lower-chromosphere brightening. 

The exposure time of observations using the 1600 \,\AA\ filtergram of AIA/SDO is almost constant ($\sim$3 s) from 2010-2015; thus, we used the original photon 
numbers for feature calculations. We set a threshold intensity to determine the brightening area as 700 photons cm$^{-5}$ s$^{-1}$ pix$^{-1}$. The threshold intensity 
was determined for features to show large variations by a parameter survey. The total intensity of the UV brightening was calculated by integrating the intensity above 
the threshold over the pixels of the determined brightening area.
 
We also used GOES X-ray data in the range of 1-8 \,\AA\, as an indicator of previous and current flare activities proposed by several authors \citep[e.g.,][]{zir91, whe04}. 
We measured the background level of the X-ray intensity by averaging the light curve of X-ray emission over 1 h and 4 h. We derived the maximum intensity one day before 
an image and counted the number of previous flares in an AR one day before and for the total period after the AR emergence, referred to as the 1-day history and the total 
history of X/M-class flares, respectively. 

\clearpage
%
%
%
\subsection{Classification by Machine Learning}

We used three machine-learning algorithms for comparison: the support vector machine (SVM), k-nearest neighbors (k-NN), and extremely randomized trees 
(ERT). Each algorithm was used as a classifier of the flare class and optimized to maximize the skill score (TSS, explained in a later section).

\subsubsection{Support Vector Machine (SVM) Classifier }

The SVM is a pattern recognition model using supervised learning \citep{vap63, bos92, cor95}. It is an algorithm of classifiers that uses a linear input to determine 
the maximum-margin hyperplane with the largest margin relative to certain points that belong to each group of the training sample. The learning process involves 
solving an optimization problem using Lagrange multipliers and the KKT condition. When the number of training samples increases, the calculation time rapidly 
increases. Here, we used a radial basis function kernel (RBF kernel or Gaussian kernel). The hyperparameter $C$ was set to $C$=10, which gave the most promising 
results. Following the standard method, the other hyperparameter $\gamma$ was set to $\gamma$=1/(the number of features).

\subsubsection{k-Nearest Neighbor (k-NN) Classifier}

The k-NN algorithm is a classifier based on the nearest instances in a feature space and is the simplest machine-learning algorithm in this study \citep{das91}. 
The classification of objects is determined by the votes of the nearest groups, that is, an object is assigned to the most popular class of the nearest $k$ objects, 
where $k$ is an integer. When $k$=1, an object is classified as being the same as the nearest object. Each feature is described by a position vector, and the 
distances among features are measured by the Euclidean or Manhattan distance. The k-NN algorithm is likely to be affected by the locality of the data. The 
selection of $k$ is performed by several heuristics. A large $k$ can reduce noise but make the border obscure. In this paper, to optimize our model, we set 
$k$=1, for which the nearest instance in the training dataset defines the prediction. Furthermore, we adopted the Manhattan distance, i.e., the distance 
$d_1({\bf x}, {\bf y}) = {\displaystyle \sum_{k=1}^n} |{\bf x}_k - {\bf y}_k|$.

\subsubsection{Extremely Randomized Trees (ERT) Classifier}

The random forest \citep{bre01} fits the number of decision-tree classifiers on various subsamples of a dataset and uses averaging to improve the prediction 
accuracy. It selects random splits to separate the subset of a node into two subsets for each of the randomly selected features, and the best split is chosen. 
In the ERT classifier \citep{geu06}, a random subset of candidate features is used, but thresholds are selected at random for each candidate feature and the best 
of these randomly generated thresholds is chosen. ERT prevents overfitting and increases the calculation speed by parallelization. We set the number of trees 
to 300 in this paper.

Another advantage of ERT is the possibility of calculation of the importance of features. Breiman (2001) proposed a method of evaluating the importance of a 
feature $X_m$ for predicting $Y$ in a tree structure $T$ by adding the decreases in the weighted impurity $p(t)\Delta$ $i$($s_t$, $t$) for all nodes $t$ where 
$X_m$ is used, then averaging over all $N_T$ trees in the forest:
\begin{equation}
Imp(X_m) = \frac{1}{N_T} \sum_{T} \sum_{t \in T: v(s_t) = X_m} p(t)\Delta i(s_t, t)
\end{equation}
where $p(t)$ is the proportion $N_t$/$N$ of samples reaching node $t$ and $v$($s_t$) is the feature used in split $s_t$ \citep[see also][]{lou13}. The decrease 
in some impurity measure $i$($t$) (e.g., the Gini index, the Shannon entropy, or the variance of $Y$) at node $t$ is defined by the following formula:
\begin{equation} 
\Delta i(s, t) = i(t) - p_L i(t_L) - p_R i(t_R),
\end{equation}
where $p_L$=$N_{t_L}$/$N$ and $p_R$=$N_{t_R}$/$N$, and the split $s_t$=$s$* for which the partition of the $N$ node samples into two subsets $t_L$ and $t_R$ 
maximizes the decrease in the impurity is identified. When nodes become pure in terms of $Y$, the construction of the tree stops. We use Gini index as the 
impurity function, and this measure is known as the Gini importance or the mean decrease Gini.

\clearpage
%
%
%
\subsection{Standardization, Evaluation, and Cross-Validation}
\subsubsection{Standardization}

The extracted solar features have different units and different scales; thus, data standardization is required. The standardization strongly affects the prediction 
accuracy, although this has not been widely acknowledged by the solar flare forecast community. We used the Z-value for standardization, i.e.,
\begin{equation}
Z = (X - \mu )/\sigma,
\end{equation}
where $X$ is the original value of the extracted solar feature, $\mu$ is the mean, and $\sigma$ is the standard deviation \citep[e.g.][]{bis06}. Therefore, Z-values 
are expressed in terms of standard deviations from means. As a result, these Z-values have a distribution with a mean of 0 and a standard deviation of 1. For parameters 
with a large-scale variation, we took the logarithm first and then calculated the Z-value.

\subsubsection{Redistribution to Training/Testing Dataset}

The solar feature database with 1 h cadence was fully shuffled and randomly separated in two datasets with a size ratio of 7 to 3 to obtain the datasets for training and 
testing, respectively. This ratio of 7 to 3 was adopted from previous works \citep{ahm13, bob15}. Note that the solar feature database is appended with flare labels when the 
sample is within 24 h of an X/M-class flare. Thus, there are at most 24 positive events per flare, and these events are included in both the training and testing databases.

\subsubsection{Validation by TSS and Cross-Validation}

We evaluated the prediction results using the past data in 2010-2015 with a skill score, the true skill statistic (TSS). This is also called the Hanssen-Kuiper skill score or 
Peirce skill score, and is defined by
\begin{equation}
TSS = \frac{TP}{TP + FN} - \frac{FP}{FP + TN},
\end{equation}
where TP, FN, FP, and TN are the numbers of true positives, false negatives, false positives, and true negatives, respectively. The score has a range of -1 to +1, with 0 
representing no skill and 1 representing perfect prediction. The TSS expresses the hit rate relative to the false alarm rate, and it remains positive provided the hit rate is 
greater than the false alarm rate. Flare prediction is an imbalanced problem, which means that negative events are much more frequent than positive ones. Bloomfield et al. 
(2012) suggested the use of the TSS because it is not affected by imbalanced problems \citep[see also][]{bob15}. This is why we selected the TSS for the evaluation of our 
prediction results.

It is important to show that our model did not suffer from overfitting. To show the validity of our approach, we used cross-validation (CV), which is the standard approach in this 
field. There are several types of CV approach such as K-fold CV, shuffle and split CV, leave-one-out CV, and so forth. In K-fold CV, the dataset is partitioned into K partitions 
and one partition acts as the validation set, and K=5 or K=10 is usually used. Note that the validation set acts as a test set, but technically it is called the validation set.

Since there are much fewer positive samples than negative samples in solar-flare classification, a large validation set contains more positive samples. This allows us to analyze 
the common features of misclassification results compared with the case where a smaller validation set is used, such as in 10-fold CV.

For the above reason, we selected shuffle and split CV to show the validity of our model. The dataset was shuffled and partitioned into training and validation sets. The size 
ratio of the two sets was 7:3, which is widely used in machine-learning and data-mining studies. This process was executed 10 times and the average results are shown.

\clearpage
%
%
%
\section{Prediction Results}

We performed supervised machine learning using the solar feature database to predict the maximum class of flares occurring in the following 24 h. We used three 
machine-learning algorithms, k-NN, SVM, and ERT, to reveal which algorithm is the most effective for a flare prediction model. We predicted two types of flare: X-class 
flares and $\ge$M-class flares. The prediction results for the three algorithms are summarized in Table 3, where in addition to a contingency table, the average TSS 
and the standard deviation as the error calculated by 10-times shuffle and split CV are listed.

The three algorithms show different prediction performances. Here, the feature database we used includes the previous flare activity derived from GOES data. For 
the prediction of X-class flares, the TSS was 0.91$\pm$0.03 for k-NN, 0.88$\pm$0.03 for SVM and 0.82$\pm$0.04 for ERT. For $\ge$M-class flares, the TSS was 
0.912$\pm$0.005 for k-NN, 0.870$\pm$0.007 for SVM, and 0.71$\pm$0.02 for ERT. Consequently, the k-NN algorithm was found to show the highest performance 
among the three algorithms on the TSS, both for X-class and $\ge$M-class flare prediction, followed by SVM then ERT. However, the FP was smallest for ERT. 
In Table 3, since the error of TSS is sufficiently small, the overfitting is small.

Table 4 shows the prediction results when the flare history for the previous day, the flare history during the whole period after the appearance of an AR, and the 
maximum X-ray intensity for the previous day are neglected. This result shows only the contribution of magnetogram and chromospheric (UV) images to flare prediction. 
For the prediction of X-class flares, the TSS was 0.91$\pm$0.02 for k-NN, 0.86$\pm$0.02 for SVM, and 0.62$\pm$0.03 for ERT, and for $\ge$M-class flare prediction, 
the TSS was 0.904$\pm$0.005 for k-NN, 0.856$\pm$0.009 for SVM, and 0.63$\pm$0.01 for ERT. By comparing Tables 3 and 4, we found that the TSS values in Table 4 
are within the error of the values reported in Table 3 for k-NN and SVM, so there is no statistical difference between the two. Only for ERT, we found a slight increase 
in the TSS upon consideration of the previous flare activity.

The importance of features calculated by ERT is also given in Table 1 for the case where the previous flare activity is considered. According to Table 1, the most 
effective feature for X-class flare prediction is the total history of X/M-class flares in an AR, followed by the maximum X-ray intensity one day before an image and 
the 1-day history of X/M-class flares. The next most effective features are the total length of magnetic neutral lines, the number of neutral lines, the unsigned 
magnetic flux, and the UV brightening area. The average magnetic field is ranked next, followed by features derived from the vector magnetogram and the time 
derivative of each feature over 24 h.

We included novel features, such as UV brightening and the time derivative of features over 24 h, 12 h and 2 h. For UV brightening, the importance of the brightening 
area and the total intensity is relatively high, while the importance of the maximum intensity is very low. When we compare the time derivatives over different time 
scales, the time derivative over 24 h is effective for flare prediction, but those over 12 h and 2 h are ineffective. Note that the magnetic free energy, the shear angle, 
and the unsigned magnetic flux near magnetic neutral lines have not been considered in this paper.

Furthermore, we compared the TSS for different detection areas of ARs, including or excluding near-limb regions. We set the detection area as the full disk, 
an intermediate area within $\pm$53$^{\circ}$ of the CM, and the disk center with a focusing area within $\pm$37$^{\circ}$. The prediction results in the latter 
two cases are summarized in Table 5. For X-class flare prediction in the intermediate case, the TSS was 0.92$\pm$0.03 for k-NN, 0.89$\pm$0.02 for SVM and 
0.88$\pm$0.03 for ERT. For the disk center, the TSS was 0.94$\pm$0.02 for k-NN, 0.92$\pm$0.03 for SVM and 0.88$\pm$0.06 for ERT. ERT is greatly improved 
by neglecting the near-limb ARs. Consequently, when the near-limb regions were neglected, the TSS was improved for all the algorithms. However, we also stress 
that in an actual operational setting, the TSS with consideration of the near-limb regions is more realistic.

\clearpage
%
%
%
\section{Summary and Discussion}

We developed a flare prediction model with the supervised machine-learning techniques using solar observations of a vector magnetogram and UV brightening. 
By detecting ARs, we extracted novel features and attached flare labels. Using training and test datasets constructed from the fully shuffled dataset, we performed 
machine learning to predict the maximum class of flares that occur in the following 24 h after observation images. An aim of this paper was to reveal which machine-learning 
algorithm is most suitable for a flare prediction model, and we compared three algorithms for the first time. Ranking of the importance of our novel features was another 
aim of this paper, and we attempted to compare the effectiveness of different features for flare prediction. 

Our prediction model achieved a skill score, the TSS, of greater than 0.9. The average performance of the k-NN algorithm was superior to those of SVM and ERT. 
One of the reasons why the TSS is improved with our model is the use of standardization, which strongly affects the prediction accuracy, although this has not 
been widely acknowledged by the solar flare forecast community. Here we used the Z-value for standardization. Furthermore, the optimization of our model such 
as by incorporating the Manhattan distance, improved the TSS.

In the daily forecast operations at NICT space weather forecast center, which use the knowledge of experts, the TSS was 0.21 for X-class flares and 0.50 for 
$\ge$M-class flares during the period 2000-2015 \citep{kub16}. At the Solar Influences Data Center (SIDC) of the Royal Observatory of Belgium, the TSS was 
0.34 for $\ge$M-class flares during the period 2004-2012 \citep{dev14}. Thus, our prediction model appears to achieve better performance than human operations. 
On the other hand, using with the fully shuffled dataset, several positive events before a flare can be divided into both training and test datasets, and consequently 
the prediction score is increased. In particular, k-NN was most effective in this study and gave the highest TSS, similarly to in other studies.

We also found that the TSS varies with the detection area; we considered the full-disk area including the near-limb regions, an intermediate area, and the disk-center 
focusing area. Upon neglecting the near-limb data, the accuracy of the features extracted from observation datasets was increased, thus improving the TSS. On 
the other hand, in an operational setting, a dataset with the near-limb region is more realistic, and the evaluation would be more similar to the human operations.

Next, we investigated the ranking of the importance of features. We showed that the previous flare activity such as the flare history in an AR and the maximum 
X-ray intensity in the previous day are the most important. The configurations of magnetic neutral lines, the unsigned magnetic flux, and the area of UV brightening 
are next most important. We also showed that the time derivative of features over 24 h is useful for prediction, while the time derivatives over 12 h and 2 h are not. 
We also found that the features of the vector magnetogram have only moderate importance, although our model did not include the magnetic free energy and the 
shear angle of magnetic fields to the magnetic neutral line.

The importance of the previous flare activity has been pointed out by several authors. The tendency for regions that have already flared to soon flare again is 
referred to as persistence in the flare forecast literature \citep{zir91}. Wheatland (2004) pointed out that future flare prediction is improved by adding the history 
of the occurrence of flares (of all sizes) to the McIntosh classification model by using a Bayesian approach. Welsch et al. (2009) showed that the flare flux averaged 
over a 24 h window exhibits some discriminant power by calculating the discriminant function coefficient \citep[e.g.][]{bar07, lek07, lek03}. However, the relative 
importance of previous flare activity was not shown in the previous papers, and in this paper, we directly showed that it is an important indicator for future flare 
prediction.

Welsch et al. (2009) also showed that the proxy Poynting flux and the unsigned flux around strong magnetic neutral lines (R-value) are important indicators of flares 
because they are related to the dynamics of flux emergence and are recognized as flare triggers. These features are not included in our study, but instead the 
maximum length, the total length, and the number of magnetic neutral lines in an AR were found to have high importance. The activity of flux emergence is also 
correlated with chromospheric brightening, which we adopted for the first time. The brightening is mainly observed along magnetic neutral lines, and it was found 
that the area of chromospheric brightening is a useful indicator of flare prediction, while the maximum intensity of the brightening is less useful. The mechanisms 
of heating and emission in the chromosphere are not so simple, suggesting that the chromospheric intensity may be a less useful indicator.

The amount of flux emergence can also be measured as the time differential of magnetic flux near magnetic neutral lines. Welsch et al. (2009) differentiated 
the magnetic flux over 90 min and concluded that the 90 min differential is too short to be a good indicator of flare prediction. Studies by Schrijver et al. (2005) 
and Longcope et al. (2005) suggest that the timescale for coronal relaxation via flaring and reconnection is on the order of 24 h. This is because the magnetic 
configuration is changed by flux emergence on the order of 24 h, not on the order of 12 h and 2 h. Furthermore, the magnetic configuration varies over a short 
time scale but only in a local area. Therefore, the time differential of features without averaging over the whole area of an AR is better for predicting flares.

Finally, the prediction score greatly depends on the datasets used for training and testing and how the database is separated into two for the training and testing. 
Separating the data into years, for example, using the data for 2010-2013 for training and the data for 2014-2015 for testing, markedly decreased the prediction 
score. This is because the samples in the two datasets were completely unrelated to each other and no similar positive events were included in both sets of data 
for training and testing, leading to a more severe condition for prediction than that in the case of fully randomly shuffled datasets. As a future work, we intend to 
examine the dependence of the prediction score on the dataset and to search for the optimal operational setting.

\acknowledgments
We acknowledge Dr. K. D. Leka for her useful comments and suggestions. We also acknowledge the referee for his/her great effort to review our paper and giving 
us useful comments. This work is supported by KAKENHI grant Number JP15K17620. The data used here are courtesy of NASA/SDO and the HMI science team, 
as well as the Geostationary Satellite System (GOES) team.
\\

\begin{deluxetable}{ l  l  l c }
\tabletypesize{\footnotesize}
\tablecaption{The extracted solar features and the importance.\label{tbl1}}
\tablewidth{0pt}
\startdata
\tableline
\tableline
Number	& \colhead{Features} & \colhead{Description} & \colhead{Importance} \\
\tableline
1   & Xhis                         & Total history of X-class flares in an AR             & 0.0519 \\[+0.0cm]
2   & Xmax1d                    & Maximum X-ray intensity one day before           & 0.0495 \\[+0.0cm]
3   & Mhis                        & Total history of M-class flares in an AR             & 0.0365 \\[+0.0cm]
4   & TotNL                      & Total length of magnetic neutral lines in an AR   & 0.0351 \\[+0.0cm]
5   & Mhis1d                     & 1-day history of M-class flares                         & 0.0342 \\[+0.0cm]
6   & NumNL                     & Number of magnetic neutral lines                      & 0.0341 \\[+0.0cm]
7   & USFlux$^\dagger$      & Total unsigned flux                                          & 0.0332 \\[+0.0cm]
8   & CHArea                    & Chromospheric (UV) brightening area   & 0.0235 \\[+0.0cm]
9   & Bave                        & Averaged magnetic field                                            & 0.0230 \\[+0.0cm]
10 & Xhis1d                      & 1-day history of X-class flares                                  & 0.0224 \\[+0.0cm]
11  & TotBSQ$^\dagger$    & Total magnitude of Lorentz force                      & 0.0199 \\[+0.0cm]
12  & Bmax                       & Maximum magnetic field                                   & 0.0193 \\[+0.0cm]
13  & MeanGAM$^\dagger$ & Mean angle of the field from the radial direction  & 0.0179 \\[+0.0cm]
14  & dt24SavNCPP           & Time derivative of SavNCPP over 24 h              & 0.0171 \\[+0.0cm]
15  & dt24TotNL                & Time derivative of TotNL over 24 h                  & 0.0169 \\[+0.0cm]
16  & dt24TotBSQ             & Time derivative of TotBSQ over 24 h                & 0.0164 \\[+0.0cm]
17  & TotFz$^\dagger$       & Sum of Z-component of Lorentz force              & 0.0160 \\[+0.0cm]
18  & dt24TotFY               &  Time derivative of TotFY over 24 h                  & 0.0156 \\[+0.0cm]
19  & Area$^\dagger$        & Area of the strong field in an AR                        & 0.0153 \\[+0.0cm]
20  & TotFY$^\dagger$      & Sum of Y-component of Lorentz force              & 0.0152 \\[+0.0cm]
21  & dt24TotFX               & Time derivative of TotFX over 24 h                  & 0.0152 \\[+0.0cm]
22  & SavNCPP$^\dagger$ & Modules of the net current per polarity             & 0.0150 \\[+0.0cm]
23  & TotUSJz$^\dagger$  & Total unsigned vertical current                         & 0.0149 \\[+0.0cm]
24  & dt24TotFZ               & Time derivative of TotFz over 24 h                   & 0.0145 \\[+0.0cm]
25  & MeanJzh$^\dagger$  & Mean current helicity (Bz contributions)            & 0.0144 \\[+0.0cm]
26  & ABSnJzh$^\dagger$  & Absolute value of the net current per polarity    & 0.0137 \\[+0.0cm]
27  & CHAll                      & Total chromospheric (UV) brightening    & 0.0134 \\[+0.0cm]
28  & TotFx$^\dagger$      & Sum of X-component of Lorentz force              & 0.0132 \\[+0.0cm]
29  & dt24USflux               & Time derivative of USflux over 24 h                 & 0.0131 \\[+0.0cm]
30  & TotUSJh$^\dagger$  & Total unsigned current helicity                        & 0.0129 \\[+0.0cm]
31  & dt24Area                 & Time derivative of Area over 24 h              & 0.0128 \\[+0.0cm]
32  & MeanGBt$^\dagger$ & Mean gradient of the total field                  & 0.0125 \\[+0.0cm]
33  & Max dxBz                 & Maximum of dBz/dx                                  & 0.0116 \\[+0.0cm]
34  & dt24ABSnJzh           & Time derivative of ABSnJzh over 24 h        & 0.0115 \\[+0.0cm]
35  & Max dyBz                 & Maximum of dBz/dy                                  & 0.0112 \\[+0.0cm]
36  & MeanGBz$^\dagger$ & Mean gradient of the vertical field              & 0.0112 \\[+0.0cm]
37  & MeanJzd$^\dagger$  & Mean vertical current density                    & 0.0111 \\[+0.0cm]
38  & dt12Area                 & Time derivative of Area over 12 h               & 0.0110 \\[+0.0cm]
39  & dt24TotUSJz           & Time derivative of TotUSJz over 24 h         & 0.0110 \\[+0.0cm]
40  & dt24Bmax                & Time derivative of Bmax over 24 h              & 0.0107 \\[+0.0cm]
41  & MaxNL                    & Maximum length of magnetic neutral lines    & 0.0107 \\[+0.0cm]
42  & Xflux4h                   & Averaged X-ray flux over 4 h                      & 0.0106 \\[+0.0cm]
43  & dt24CHArea             & Time derivative of CHArea over 24 h           & 0.0103 \\[+0.0cm]
44  & dt12Bmax                & Time derivative of Bmax over 12 h              & 0.0097 \\[+0.0cm]
45  & MeanGBh$^\dagger$ & Mean gradient of the horizontal field           & 0.0092 \\[+0.0cm]
46  & Xflux1h                   & Averaged X-ray flux over 1 h                      & 0.0091 \\[+0.0cm]
47  & dt12USflux              & Time derivative of USflux over 12 h             & 0.0090 \\[+0.0cm]
48  & dt24 Max graB         & Time derivative of Max. grad. Bz over 24 h    & 0.0088 \\[+0.0cm]
49  & dt24 Max dzBy         & Time derivative of Max. dBy/dz over 24 h     & 0.0088 \\[+0.0cm]
50  & dt24 TotUSJh         & Time derivative of TotUSJh over 24 h          & 0.0081 \\[+0.0cm]
51  & dt24 NumNL       & Time derivative of NumNL over 24 h                    & 0.0079 \\[+0.0cm]
52  & dt24 MaxdxBz     & Time derivative of MaxdxBz over 24 h                  & 0.0079 \\[+0.0cm]
53  & dt24MeanJzh      & Time derivative of MeanJzh over 24 h                 & 0.0078 \\[+0.0cm]
54  & dt24MaxNL         & Time derivative of MaxNL over 24 h                     & 0.0075 \\[+0.0cm]
55  & dt02 Area           & Time derivative of Area over 2 h                         & 0.0071 \\[+0.0cm]
56  & dt24 CHAll          & Time derivative of CHAll over 24 h                      & 0.0071 \\[+0.0cm]
57  & Bmin                  & Minimum magnetic field of Bz                             & 0.0071 \\[+0.0cm]
58  & CHMax               & Maximum intensity of chromospheric (UV) brightening  & 0.0062 \\[+0.0cm]
59  & dt02 Bmax          & Time derivative of Bmax over 2 h                        & 0.0061 \\[+0.0cm]
60  & dt24 CHMax        & Time derivative of CHMax over 24 h                    & 0.0049 \\[+0.0cm]
61  & dt24 MeanGBz    & Time derivative of MeanGBz over 24 h           & 0.0028 \\[+0.0cm]
62  & dt24 MeanGBh    & Time derivative of MeanGBh over 24 h          & 0.0021 \\[+0.0cm]
63  & dt24 MeanGBt    & Time derivative of MeanGBt over 24 h           & 0.0002 \\[+0.0cm]
64  & dt24MeanGAM    & Time derivative of MeanGAM over 24 h          & 0.0002 \\[+0.0cm]
65  & dt24MeanJzd      & Time derivative of MeanJzd over 24 h            & 0.0000 \\[+0.0cm]
%
\enddata
\tablecomments{The formulae of the features attached with $\dagger$ marks are shown in Table 2 and in Bobra \& 
Couvidat (2015). The importance was calculated by ERT for X-class flare prediction.}
\end{deluxetable}


\begin{deluxetable}{ l  l  c }
\tabletypesize{\footnotesize}
\tablecaption{Formulae of AR features \label{tbl10}}
\tablewidth{0pt}
\startdata
\tableline
\tableline
Keyword & \colhead{Description} & \colhead{Formula} \\
\tableline
{\scriptsize TOTUSJH} & Total unsigned current helicity & $H_{c_{total}} \propto \sum |B_z \cdot J_z|$  \\[+0.10cm]
{\scriptsize TOTBSQ}  & Total magnitude of Lorenz force & $F \propto \sum B^2$  \\[+0.10cm]
{\scriptsize TOTUSJZ} & Total unsigned vertical current & $J_{z_{total}} = \sum |J_z| dA$  \\[+0.10cm]
{\scriptsize ABSNJZH} & Absolute value of the net current per polarity & $H_{c_{abs}} \propto |\sum B_z \cdot J_z|$  \\[+0.10cm]
{\scriptsize SAVNCPP} & Sum of the modules of the net current per polarity & $J_{z_{sum}} \propto |\displaystyle \sum^{B_z^{+}} J_z dA| + |\displaystyle \sum^{B_z^{-}} J_z dA|$  \\[+0.10cm]
{\scriptsize USFLUX}    & Total unsigned flux & $\Phi = \sum |B_z| dA$  \\[+0.10cm]
{\scriptsize AREA-ACR} & Area of strong field pixels in the active region & $Area = \sum Pixels$  \\[+0.10cm]
{\scriptsize TOTFZ}      & Sum of z-component of Lorenz force & $F_z \propto \sum (B_x^2 + B_y^2 -B_z^2) dA$  \\[+0.10cm]
%
{\scriptsize EPSZ}        & Sum of z-component of normalized Lorentz force & $\delta F_z \propto \frac{\sum (B_x^2 + B_y^2 -B_z^2)}{\sum B^2}$  \\[+0.10cm]
{\scriptsize MEANGAM}  & Mean angle of field from radial & $\overline{\gamma} = \frac{1}{N}\sum arctan \left( \frac{B_h}{B_z} \right)$ \\[+0.10cm]
{\scriptsize MEANGBT}  & Mean gradient of total field & $\overline{|\nabla B_{tot}|} =\frac{1}{N} \sum \sqrt{ \left( \frac{\partial B}{\partial x} \right)^2 + \left( \frac{\partial B}{\partial y} \right)^2 }$  \\[+0.10cm]
{\scriptsize MEANGBZ} & Mean gradient of vertical field & $\overline{|\nabla B_z|} =\frac{1}{N}\sum \sqrt{\left( \frac{\partial B_z}{\partial x} \right)^2 + \left( \frac{\partial B_z}{\partial y} \right)^2 }$  \\[+0.10cm]
{\scriptsize MEANGBH} & Mean gradient of horizontal field & $\overline{|\nabla B_h|} =\frac{1}{N}\sum \sqrt{\left( \frac{\partial B_h}{\partial x} \right)^2 + \left( \frac{\partial B_h}{\partial y} \right)^2 }$ \\[+0.10cm]
{\scriptsize MEANJZH}  & Mean current helicity ($B_z$ contribution) & $\overline{H_c} \propto \frac{1}{N} \sum B_z \cdot J_z$  \\[+0.10cm]
{\scriptsize TOTFY}      & Sum of y-component of Lorentz force & $F_y \propto \sum B_y B_z dA$  \\[+0.10cm]
{\scriptsize MEANJZD} & Mean vertical current density & $\overline{J_z} \propto \frac{1}{N} \sum \left( \frac{\partial B_y}{\partial x} - \frac{\partial B_x}{\partial y} \right) $  \\[+0.10cm]
{\scriptsize TOTFX}     & Sum of x-component of Lorentz force & $F_x \propto -\sum B_x B_z dA$  \\[+0.10cm]
{\scriptsize EPSY}       & Sum of y-component of normalized Lorentz force & $\delta F_y \propto \frac{-\sum B_y B_z}{\sum B^2}$  \\[+0.10cm]
{\scriptsize EPSX}       & Sum of x-component of normalized Lorentz force & $\delta F_x \propto \frac{\sum B_x B_z}{\sum B^2} $  \\[+0.10cm]
%
\enddata
\tablecomments{The formulae in this table are quoted from Bobra \& Couvidat (2015).}
\end{deluxetable}


\begin{deluxetable}{ c l  l  l  l  l  l  c }
\tabletypesize{\footnotesize}
\tablecaption{The prediction results of X-class flares and $\ge$M-class flares. \label{tbl3}}
\tablewidth{0pt}
\startdata
\tableline
\tableline
\colhead{Algorithm} & \colhead{TP} & \colhead{FP} & \colhead{FN} & \colhead{TN} & \colhead{TSS} \\
\tableline
(a) X-class flares & & & & & \\
k-NN  & 152   & 14  & 11  & 54439  & 0.91 $\pm$ 0.03 \\[+0.0cm]
SVM   & 120   & 22  & 16  & 54458  & 0.88 $\pm$ 0.03 \\[+0.0cm]
ERT    & 134   & 7   &  29 & 54446  & 0.82 $\pm$ 0.04 \\[+0.0cm]
 & & & & & & \\
(b) $\ge$M-class flares & & & & & \\	
k-NN	  &  1544  &  121  &  155  &  52796  &  0.912 $\pm$ 0.005 \\[+0.0cm]
SVM	  &  1496  &  473  &  203  &  52444  &  0.870 $\pm$ 0.007 \\[+0.0cm]
ERT    &  1216  &    39  &  483  &  52878  &  0.71 $\pm$ 0.02 \\[+0.0cm]
%
%
\enddata
\tablecomments{The contingency tables of prediction results of X-class flares and $\ge$M-class flares, for the three machine-learning 
algorithms, k-NN, SVM and ERT.}
\end{deluxetable}

\begin{deluxetable}{ c l  l  l  l  l  l  c }
\tabletypesize{\footnotesize}
\tablecaption{The prediction results of X-class flares and $\ge$M-class flares, neglecting features of previous flare activities. \label{tbl4}}
\tablewidth{0pt}
\startdata
\tableline
\tableline
\colhead{Algorithm} & \colhead{TP} & \colhead{FP} & \colhead{FN} & \colhead{TN} & \colhead{TSS} \\
\tableline
(a) X-class flares & & & & & \\
k-NN  & 136   & 16  & 15  & 54449  & 0.91 $\pm$ 0.02 \\[+0.0cm]
SVM   & 130   & 23  & 21  & 54442  & 0.86 $\pm$ 0.02 \\[+0.0cm]
ERT    & 87    & 4   &  49  & 54476  & 0.62 $\pm$ 0.03 \\[+0.0cm]
 & & & & & & \\
(b) $\ge$M-class flares & & & & & \\	
k-NN	  &  1570  &  173  &  167  &  52706  &  0.904 $\pm$ 0.005 \\[+0.0cm]
SVM	  &  1501  &  759  &  236  &  52120  &  0.856 $\pm$ 0.009 \\[+0.0cm]
ERT    &  1105  &    35  &  632  &  52844  &  0.63 $\pm$ 0.01 \\[+0.0cm]
%
%
\enddata
\tablecomments{The contingency tables of prediction results of X-class flares and $\ge$M-class flares, for the three machine-learning 
algorithms, k-NN, SVM and ERT.}
\end{deluxetable}

\begin{deluxetable}{ c l  l  l  l  l  l  c }
\tabletypesize{\footnotesize}
\tablecaption{The prediction results of X-class flares with different detection regions. \label{tbl5}}
\tablewidth{0pt}
\startdata
\tableline
\tableline
\colhead{Algorithm} & \colhead{TP} & \colhead{FP} & \colhead{FN} & \colhead{TN} & \colhead{TSS} \\
\tableline
(a) An intermediate area & & & & & \\
k-NN  & 87   & 8  & 5  & 43277  & 0.92 $\pm$ 0.03 \\[+0.0cm]
SVM   & 84  & 12 & 7  & 43274  & 0.89 $\pm$ 0.02 \\[+0.0cm]
ERT   & 80   & 0  & 10 & 43287  & 0.88 $\pm$ 0.03 \\[+0.0cm]
 & & & & & & \\
(b) The disk center focusing area & & & & & \\	
k-NN	  &  54  &  2  &  4  &  26782  &  0.94 $\pm$ 0.02 \\[+0.0cm]
SVM	  &  57  &  3  &  5  &  26777  &  0.92 $\pm$ 0.03 \\[+0.0cm]
ERT    &  55  &  2  &  8  &  26777  &  0.88 $\pm$ 0.06 \\[+0.0cm]
%
%
\enddata
\tablecomments{The contingency tables of prediction results of X-class flares with different detection regions: with an intermediate area within $\pm$57$^{\circ}$ of the CM 
(within 4/5 of the solar radius) and with the disk center with a focusing area within $\pm$37$^{\circ}$ (within 3/5 of the solar radius). We used the three machine-learning 
algorithms, k-NN, SVM and ERT.}
\end{deluxetable}


\clearpage

\clearpage

%
%
%
\end{document}